\documentclass[reprint,pre,floats,floatfix,aps,amsmath,amssymb]{revtex4-2}
\usepackage{graphicx}
\DeclareMathOperator\artanh{artanh\,}
\begin{document}
\title{Effective Hamiltonian approach to kinetic Ising models: Application
	to an infinitely long-range Husimi-Temperley model}
\author{V. I. Tokar}
\affiliation{G. V. Kurdyumov Institute for Metal Physics of the
N.A.S.\ of Ukraine, 36 Acad. Vernadsky Boulevard, UA-03142 Kyiv, Ukraine}
\affiliation{Universit{\'e} de Strasbourg, CNRS, IPCMS, UMR 7504,
F-67000 Strasbourg, France}
\date{\today}
\begin{abstract}
	The probability distribution (PD) of spin configurations in 
	kinetic Ising models has been cast in the form of the canonical
	Boltzmann PD with a time-dependent effective Hamiltonian (EH). It
	has been argued that in systems with extensive energy EH depends
	linearly on the number of spins $N$ leading to the exponential
	dependence of PD on the system size. In macroscopic systems the
	argument of the exponential function may reach values of the
	order of the Avogadro number which is impossible to deal with
	computationally, thus making unusable the linear master equation
	(ME) governing the PD evolution.  To overcome the difficulty,
	it has been suggested to use instead the nonlinear ME (NLME)
	for the EH density per spin. It has been shown that in spatially
	homogeneous systems NLME contains only terms of order unity even 
	in the thermodynamic limit.
	
	The approach has been illustrated with the kinetic
	Husimi-Temperley model (HTM) evolving under the Glauber
	dynamics. At finite $ N $ the known numerical results has been
	reproduced and extended to broader parameter ranges. In the
	thermodynamic limit an exact nonlinear partial differential
	equation of the Hamilton-Jacobi type for EH has been derived. It
	has been shown that the average magnetization in HTM evolves
	according to the conventional kinetic mean field equation.
\end{abstract}
\maketitle
\section{Introduction}
The equilibrium statistics of Ising-type models at a fixed temperature
can be described by the canonical probability distribution (CPD)
\cite{Huang_1987}. The latter depends only on the spin configuration
energy which is conventionally called the Hamiltonian. CPD is proportional
to the Boltzmann factor which is the exponential function of the
Hamiltonian divided by $-k_BT$---the absolute temperature in energy units
with the minus sign.  In models with short-range interactions the energy
scales linearly with the system size $N$. So if exact analytic solution is
unknown, the straightforward use of CPD in approximate calculations would
necessitate numerical calculation of the exponential function with the
argument scaling as  $ O(N) $ which can be difficult or even impossible
for large $ N $ that are of main interest in statistical physics. To
deal with this problem, sophisticated combinatorial techniques have been
developed to calculate physical quantities of interest without resorting
to the numerical exponentiation of $ O(N) $ numbers and using only $ O(1)
$ quantities, such as specific magnetization and the energy density per
site \cite{englert,Huang_1987}.

From the standpoint of out of equilibrium statistics CPD is a particular
case of more general nonequilibrium probability distribution
(NPD) which coincides with CPD in thermal equilibrium. This is
conveniently formalized in the effective Hamiltonian (EH) approach
\cite{PRE96,vaks1996,Nastar2000SelfconsistentFO} where the dependence of
NPD on EH is posited to be the same as that of CPD on the equilibrium
Hamiltonian.  This is achieved by defining EH as the logarithm of NPD
multiplied by $-k_BT$.  This trick allows one to use the approximate
techniques of equilibrium statistics \cite{Huang_1987,englert} also in
the nonequilibrium case.

This, however, does not fully solve the problem because unlike
the equilibrium case where the Hamiltonian is constant in time and
supposed to be known, EH for a system out of equilibrium should be
determined separately. In the case of  kinetic Ising models
(IM) \cite{van1992stochastic,glauber,kawasaki,soft-dynamics},
EH evolves with time according to the master equation (ME) for NPD
\cite{van1992stochastic}. Because NPD depends on EH and $N$ exponentially,
the numerical solution of ME for large $ N $ encounters the same over-
and underflow difficulties as in the statistical averaging. For example,
assuming the number of spins $ N=10^4 $ the span of NPD values accessible
to calculations within quadruple precision arithmetic will be exhausted
by absolute values of EH density only slightly exceeding $ k_BT $ which
would severely restrict calculations at low temperatures. This could
be the reason why simulations in Ref.\ \cite{mori_asymptotic_2010}
were restricted to systems containing not more than 10000 spins.

In this paper we will deal with kinetic Ising-type models with
Glauber dynamics \cite{glauber} that have been used to describe
out of equilibrium kinetics in a wide variety of systems, such as
uniaxial magnets, lattice  gases, binary alloys, spin glasses,
proteins, neural networks, combinatorial optimization, etc.
\cite{Huang_1987,toga_role_2020,gouyet_description_2003,ducastelle,%
plefka_modified_2002,BPZ,aguilera_unifying_2021,Kirkpatrick1983OptimizationBS}.

The aim of the present paper is to suggest a modification of ME along
the lines of Ref.\ \cite{PRE96} such that the resulting evolution
equation contained exponential functions with $ O(1) $ arguments and
in the case of homogeneous systems the equation for EH contained only
terms of order unity. It will be shown that this can be achieved at
the cost of transforming the linear ME into a nonlinear evolution
equation (henceforth abbreviated as NLME).  Its derivation will be
given in Sec.\ \ref{nlme}. NLME for the kinetic Husimi-Temperley
model (HTM) (also known as the infinitely long-range Ising-,
the mean field- (MF), the Curie-Weiss-, and the Weiss-Ising model
\cite{weiss-ising,matkowsky1984asymptotic,chakrabarti_dynamic_1999,%
mori_asymptotic_2010,carinci2013}) evolving under the Glauber dynamics
\cite{glauber} will be derived in Sec.\ \ref{htm}.  Numerical merits
of NLME will be illustrated on the problem of decay of metastable
states in HTM. The decay problem was previously investigated in Refs.\
\cite{weiss-ising,matkowsky1984asymptotic,salinas_introduction_2001,%
mori_asymptotic_2010} in the framework of ME, the Fokker-Planck equation,
and the Monte Carlo simulations. In Ref.\  \cite{mori_asymptotic_2010} a
scaling law for the lifetime of metastable states was suggested and shown
to be valid for large absolute values of a scaling parameter. In Secs.\
\ref{numerical} and \ref{calculation} it will be shown that the use of
NLME allows one to extend the testing range of the scaling law as well
as the accuracy of the calculated lifetimes of the {metastable} states by
orders of magnitude. In Sec.\ \ref{thrm-limit} it will be shown that NLME
for HTM in the thermodynamic limit reduces to a nonlinear differential
equation of the first order. Its characteristic equations will be derived
and it will be shown that the conventional MF kinetic equation of Refs.\
\cite{suzuki-kubo,chakrabarti_dynamic_1999,gouyet_description_2003}
describes a characteristic that passes through the minimum of a free
energy function. The description of hysteresis with the use of MF
equation, however, is inconsistent because it does not predict a
vanishing hysteresis loop area at zero frequency of the oscillating
external magnetic field \cite{chakrabarti_dynamic_1999}. In Sec.\
\ref{calculation} a way of resolving this difficulty in HTM framework
will be suggested which besides purely theoretical interest may be
also of practical importance. It will be argued that HTM at finite $N$
exhibits the N\'eel-type relaxation \cite{neel1949} which is of interest
in biomedical applications where the hysteretic behavior of magnetic
nanoparticles is of major importance \cite{biomed}.  In concluding Sec.\
\ref{conclusion} a brief summary will be presented and further arguments
given to support the suggestion that NLME is a prospective approach to
the solution of stochastic models of the Ising type.
\section{\label{nlme}Effective Hamiltonian approach to kinetic Ising 
models}
For brevity, the set of $ N $ Ising spins $\sigma_i=\pm1$, $i = 1,N$,
will be denoted as $ \pmb{\sigma}=\{\sigma_i\}$.  In the stochastic
approach to nonequilibrium kinetics the statistical properties of an
out of equilibrium system can be described by the time-dependent NPD
$P(\pmb{\sigma},t)$ which satisfies ME \cite{van1992stochastic}
\begin{equation}
	{P}_t(\pmb{\sigma},t)=\sum_{\pmb{\sigma}^{\prime}}
	[r(\pmb{\sigma}^{\prime}\to\pmb{\sigma},t)
		P(\pmb{\sigma}^{\prime},t)
	-r(\pmb{\sigma}\to\pmb{\sigma}^{\prime},t)
P(\pmb{\sigma},t)].
	\label{ME}
\end{equation}
Here and in the following by subscript $t$ we will denote the partial time
derivative; the transition rates $ r $ in the kinetic IM will be chosen
according to Refs.\ \cite{glauber,kawasaki} (the ``soft'' Glauber dynamics
\cite{soft-dynamics} can be treated similarly with minor modifications)
\begin{equation}
r(\pmb{\sigma}\to\pmb{\sigma}^\prime,t)=
\frac{1}{\tau_0}
\frac{e^{{-\cal H}^0(\pmb{\sigma}^\prime,t)}}
{e^{-{\cal H}^0(\pmb{\sigma}^\prime,t)}
+e^{-{\cal H}^0(\pmb{\sigma},t)}}.
	\label{rss'}
\end{equation}
where $1/\tau_0$ is the rate of transition
$\pmb{\sigma}\to\pmb{\sigma}^\prime$ and the dimensionless
Hamiltonian ${\cal H}^0$ is assumed to include the factor
$\beta=1/k_BT$ as a parameter. ${\cal H}^0$ may depend  on
time which is necessary, for example, in modeling the hysteresis
\cite{tome_dynamic_1990,chakrabarti_dynamic_1999,zhu_hysteresis_2004}.
The dependence of the Hamiltonian on $\pmb{\sigma}$ can be arbitrary but
in this study we will assume that ${\cal H}^0$ is an extensive quantity,
that is, it scales linearly with the system size $N$ \cite{Huang_1987}.
In this case the exponential functions in Eq.\ (\ref{rss'}) scale with
$N$ as
\begin{equation}
	e^{-{\cal H}^0(\pmb{\sigma},t)}
	= e^{-Nu^0(\pmb{\sigma},t)},
	\label{u0}
\end{equation}
where $u^0=O(1)$ is the Hamiltonian density per spin. The exponential
behavior in Eq.\ (\ref{u0}) may hinder numerical solution of ME
(\ref{ME}) because the terms on the right hand side (rhs) of Eq.\
(\ref{rss'}) will suffer from the problem of numerical over- and
underflow at sufficiently large $N$. However, when the transition
$\pmb{\sigma}\to\pmb{\sigma}^\prime$ is local,  this difficulty is
easily overcome. Multiplying the numerator and denominator in Eq.\
(\ref{rss'}) by $\exp\{\tfrac{1}{2}[{\cal H}^0(\pmb{\sigma},t) +{\cal
H}^0(\pmb{\sigma}^\prime,t)]\}$ one arrives at
\begin{equation}
	r(\pmb{\sigma}\to\pmb{\sigma}^\prime)
	=\frac{1}{2\tau_0} \frac{\exp{[\tfrac{1}{2}\Delta^{{\cal H}^0} 
	(\pmb{\sigma},\pmb{\sigma}^\prime,t)}]}
	{\cosh[{\tfrac{1}{2}\Delta^{{\cal H}^0} (\pmb{\sigma},\pmb{\sigma}^\prime,t)}]}
	\label{r}
\end{equation}
where
\begin{equation}
	\Delta^{{\cal H}^0}(\pmb{\sigma},\pmb{\sigma}^\prime,t)=
	{\cal H}^0(\pmb{\sigma},t)-{\cal H}^0(\pmb{\sigma}^{\prime},t).
	\label{delta}
\end{equation}
In lattice models with local spin interactions the $O(N)$ scaling of
${\cal H}^0$ is a consequence of the summation over $N$ lattice sites. If
configurations $\pmb{\sigma}$ and $\pmb{\sigma}^\prime$ differ only
locally, then the difference in Eq.\ (\ref{delta}) in such models vanishes at
large distances from the  flipping spin so the summation over sites will
be spatially restricted leading to $\Delta^{{\cal H}^0}=O(1)$, hence,
all terms in Eq.\ (\ref{r}) will be of order unity.  (The locality will
be discussed in more detail below in Sec.\ \ref{local_int}.)

We note that Eq.\ (\ref{delta}) is antisymmetric under the
exchange of spin arguments so the denominator in Eq.\ (\ref{r})
is the same for both terms in  Eq.\ (\ref{ME}). Replacing it by a
constant can simplify ME while preserving its qualitative features
\cite{PRE96,chakrabarti_dynamic_1999}. In the present paper, however, more
complex Glauber prescription (\ref{rss'}) will be used for compatibility
with the majority of literature on the kinetic HTM (see, e.g., Refs.\
\cite{tome_dynamic_1990,mori_asymptotic_2010} and references therein).

Another source of the undesirable behavior briefly discussed in
the Introduction comes from NPD $P$ which should depend on $ N $
similar to CPD because it coincides with it in thermal equilibrium
\begin{equation}
	P(\pmb{\sigma},t)|_{t\to\infty}\to P^0(\pmb{\sigma})=e^{-{\cal
	H}^0(\pmb{\sigma})}/Z \label{equilibrium}
\end{equation} 
where $Z$ is the partition function and ${\cal H}^0$ in Eq.\
(\ref{equilibrium}) is assumed to be time-independent because
otherwise the equilibrium would not be attainable. It is easily
checked that $P^0$ is a stationary solution of Eq.\ (\ref{ME})
which means that at least for large $t$ NPD $P$ in Eq.\ (\ref{ME})
behaves similar to $P^0$.  This can be formalized in the EH approach
\cite{PRE96,Nastar2000SelfconsistentFO,gouyet_description_2003} where by
analogy with Eqs.\ (\ref{equilibrium}) and (\ref{u0}) NPD depends on EH $
{\cal H} $ as
\begin{equation}
	P(\pmb{\sigma},t)=e^{-{\cal H}(\pmb{\sigma},t)}\equiv 
	e^{-Nu(\pmb{\sigma},t)},
	\label{P-u}
\end{equation}
where $ {\cal H}|_{t\to\infty}\to{\cal H}^0 +\ln Z$ and $ u $ is the EH
density. In Eq.\ (\ref{P-u}) it has been assumed that the normalization
constant similar to $ \ln Z $ is included in ${\cal H}$. This is
convenient in many cases because ME (\ref{ME}) preserves normalization
so whenever possible the initial $ {\cal H}(\pmb{\sigma},t=t_0) $ is
reasonable to choose in such a way that $P(t_0)$ in Eq.\ (\ref{P-u})
was easily normalizable. This would eliminate the necessity to calculate
an equivalent of the partition function $Z$ at later stages of evolution
because in general this is a nontrivial task.

In Ref.\ \cite{PRE96} a simple way to eliminate the undesirable
exponential behavior from ME is suggested which was implemented as
follows. Dividing both sides of Eq.\ (\ref{ME}) by $ P $ from Eq.\
(\ref{P-u}) one arrives after trivial rearrangement at the evolution
equation for EH
\begin{equation}
	{\cal H}_t(\pmb{\sigma},t)=\frac{1}{\tau_0}\sum_{\pmb{\sigma}^{\prime}}
	\frac{ \exp\left(\tfrac{1}{2}\Delta^{{\cal H}^0}
-\Delta^{{\cal H}}\right)-\exp\left(-\tfrac{1}{2}
\Delta^{{\cal H}^0}\right)}{2\cosh(\tfrac{1}{2}\Delta^{{\cal H}^0})}
	\label{NME}
\end{equation}
where $\Delta^{\cal H}$ is defined as in Eq.\ (\ref{delta}) and normally
it should be an $O(1)$ quantity (see Sec.\ \ref{local_int}). The arguments
of deltas in Eq.\ (\ref{NME}) have been omitted for brevity; they are
the same as in Eq.\ (\ref{delta}). As is seen, all terms on the rhs
are $O(1)$ but because of the summation over $ \pmb{\sigma}^{\prime}
$ the equation scales linearly with $ N $. Formally, division of both
sides by $N$ would make equation $O(1)$ and fully expressed in terms of
EH density defined in Eqs.\ (\ref{u0}) and (\ref{P-u}).

Less formal results can be obtained in spatially homogeneous case.
Assuming that the sites are arranged in a Bravais lattice, that is, they
are all equivalent, an arbitrary site can be chosen as a reference point.
Next applying the cluster approach of the equilibrium alloy theory
\cite{ducastelle,Sanchez1984334} to Eq.\ (\ref{NME}) the expansion
over a complete set of orthonormal polynomials of $ \pmb{\sigma}$ can
be restricted only to those containing the chosen site.  Further, if
the interactions in the system are local  (see Sec.\ \ref{local_int}),
then the size of the system of equations to solve can be approximated by a
finite system. In practice, in the alloy thermodynamics the number of
clusters to keep in the expansion was found to be rather moderate in
many cases \cite{c_INdep_CE,WOLVERTON1997107}. The approximation can be
justified with the use of a series expansion. More specific discussion of
approximate solution of NLME is not possible because it would strongly
depend on an arbitrary initial condition and on unspecified number of
time-dependent Hamiltonian parameters.
\subsection{\label{local_int}Local interactions in EH}
Though nonequilibrium case is difficult to analyze in general terms,
in simple cases the locality can be proved.

Let us consider the case of IM with time-independent pair interactions
\begin{equation}\label{nnIsing}
	{\cal H}^{0}(\pmb{\sigma}) =
	-\tfrac{1}{2}\sum_{i,\vec{e}}K_{i,i+\vec{e}}\sigma_i
	\sigma_{i+\vec{e}}-h\sum_i\sigma_i
\end{equation} 
where $K_{ij}=J_{ij}/k_BT$, $J_{ij}$ being the pair interactions,
$h=H/k_BT$, and $H$ is the magnetic field. The second subscript of $K$
in Eq.\ (\ref{nnIsing}) is represented as the sum of $i$ and of the radius
vector $\vec{e}$ of the relative position of site $j$ with respect to $i$.

For simplicity let us consider Bravais lattices and assume that the
Hamiltonian satisfy all lattice symmetries. In this case the pair
interactions $K_{ij}$ depend only on the difference of the lattice
coordinates $i-j$ so in Eq.\ (\ref{nnIsing}) for any site $i$ only
dependence on the relative coordinate $\vec{e}$ remains.  Further, because
the absolute value of the Ising spins is unity, the absolute value of
summation over $\vec{e}$ in Eq.\ (\ref{nnIsing}) will be bounded for
all $i$ if
\begin{equation}
	\sum_{\vec{e}}\vert K(\vec{e})\vert = O(1)
	\label{abs} 
\end{equation}
which is a formal definition of the short range interaction.
The summation over $i$ in Eq.\ (\ref{nnIsing}) of $N$ $O(1)$ terms
makes the energy an extensive quantity for any spin configuration.
The criteria for multispin interactions to be short range will have
a form similar to Eq.\ (\ref{abs}) only the summation will be carried
over several radius vectors.  Though vectors $ \vec{e}$ in the summation
can be arbitrarily long, the interactions are of short range provided
the sum is convergent. Under the Glauber dynamics $\pmb{\sigma}$ and
$\pmb{\sigma}^{\prime}$ in Eq.\ (\ref{delta}) differ only by one spin
on one site, say, $i$, so only those terms in ${\cal H}^0$ in Eq.\
(\ref{nnIsing}) that contain this spin will contribute to the difference
\begin{equation}
	\Delta_i^{{\cal H}^0}=-2\sigma_i\sum_{\vec{e}}
	K(\vec{e})\sigma_{i+\vec{e}}-2h\sigma_i
	\label{delta_i}
\end{equation} 
which is $O(1)$ quantity due to Eq.\ (\ref{abs}) (here $
\pmb{\sigma}^\prime $ on the rhs in Eq.\ (\ref{delta}) has been replaced
by $ \pmb{\sigma}$ using the fact that $ \sigma_i^\prime=-\sigma_i $
while all other spins remain unchanged).

Thus, the terms in Eq.\ (\ref{NME}) with $\Delta^{{\cal H}^0}$ will cause
no problems in numerical calculations in case of short range interactions.
However, $\Delta^{{\cal H}}$ in the equation is not straightforward to
analyze because EH is not known and should be determined as the solution
of NLME.  This in general is a difficult task because the initial
condition depends in general on $2^N$ arbitrary parameters. Besides,
arbitrary time dependence in ${\cal H}^0$ is possible in the out of
equilibrium systems.  Therefore, to make the task manageable, let
us consider as an example a frequently studied problem \cite{coarsening}
of the quench from a disordered phase at high temperature $T_0$ to a
lower temperature $T$ which we assume to be also large.  In this case
one can use the high temperature expansion in powers of $\beta$ or, in
our notation, in powers of the dimensionless Hamiltonian ${\cal H}^0$.
For simplicity let us set $T_0=\infty$ which means the initial value
of EH ${\cal H}(\pmb{\sigma},t=0)=0$. Because ${\cal H}^0=O(\beta)$
and EH varies from zero at $t=0$ to ${\cal H}^0$ when $t\to\infty$,
it is also an $O(\beta)$ quantity. Omitting for simplicity the external
magnetic field ($h=0$) and expanding Eq.\ (\ref{NME}) to the first order
in $\Delta$'s it can be shown that Eq.\ (\ref{delta_i}) summed over $i$
produces $4{\cal H}^0$ and similarly for $\Delta^{\cal H}$ which, as
will be shown, has the same spin structure in this order.  Now from Eq.\
(\ref{nnIsing}) one gets to the first order in $\beta$ a linear equation
\begin{equation}
	{\cal H}_t = 2\tau_0^{-1}({\cal H}^0-{\cal H})
	\label{1st_order}
\end{equation} 
which integration gives
\begin{equation}
	{\cal H}(\pmb{\sigma},t)={\cal H}^0(\pmb{\sigma})(1-e^{-2t/\tau_0}).
	\label{H1st_order}
\end{equation}
As is seen, in this approximation EH has the same locality properties as
the physical Hamiltonian. Because of nonlinearity of Eq.\ (\ref{NME}),
higher orders in $\beta$ will be more difficult to analyze. Restricting
discussion for simplicity to the case of the nearest neighbor (nn)
interactions, it is easy to see that, e.g., in the second order in $\beta$
the squares of the first order $ {\cal H}^0 $ will introduce the pair
interactions in ${\cal H}$ of the extent not exceeding the largest
distance within the first coordination sphere.  Higher order terms in
$\beta$ will introduce even farther neighbor interactions into EH but
to any finite order they will remain bounded by some maximum extent,
thus retaining the locality when approximation to this order is adequate.

From a practical standpoint more interesting is the quench to temperatures
below $ T_c $ \cite{coarsening}.  Because the high temperature expansion
breaks down in this case, the locality of EH can be substantiated at the
initial stage of evolution within the region where expansion in time
variable $t$ converges. Assuming as above that at $t=0$ EH vanishes,
the coefficient of linear in $t$ term can be found with the use of Eq.\
(\ref{NME}) as
\begin{equation}
	{\cal H}_t\vert_{t=0}=\tau_0^{-1}\sum_i 
	\tanh(\tfrac{1}{2}\Delta_i^{{\cal H}^0}).
	\label{t-exp}
\end{equation}
Because the hyperbolic tangent is an entire function, it can be
expanded in powers of the argument at arbitrarily large values of ${\cal
H}^0\propto \beta$ at low temperatures where $ \beta $ is large. In the
case of nn interactions, only $\sigma_i$ and the nn spins will enter
the expansion which means that interactions in Eq.\ (\ref{t-exp}) will
extend not further then the diameter of the first coordination sphere. In
higher order terms of expansion in $t$ more distant interactions will
arise but they will remain short range to any finite order.

Thus, we have shown that the locality of EH holds within the regions
of convergence of two series expansions. Arguably, other expansion
techniques suitable to specific physical setups can be developed. The
convergence cannot always be guaranteed for any of the expansions but
from equilibrium statistics we know that the divergence of a series
usually signals the appearance of new physical behavior, such as the
phase transitions and/or critical phenomena. It may be expected that in
the out of equilibrium evolution the singularities in series expansions
would also reveal some new physics, such as the dynamic phase transitions
\cite{tome_dynamic_1990,chakrabarti_dynamic_1999}.

Though EH cannot be experimentally measured, the distribution of spin
configurations depends on EH exactly in the same way as the CPD depends
on ${\cal H}^0$ by EH definition. Therefore, the cluster interactions
that appear in EH during the evolution will influence the correlation
functions in exactly the same way as the physical cluster interactions
occurring in the alloy Hamiltonians   \cite{ducastelle,Sanchez1984334}.
The cluster interactions may influence such observable phenomena as the
diffusion scattering \cite{masanskii_cluster_1992} and all physical
quantities that depend on cluster statistics, such as the electrical
conductivity.  Therefore, some physical insight into out of equilibrium
alloys can be gained from the alloy studies in thermal equilibrium
\cite{ducastelle,us-ordering-potentials}.
\section{\label{htm}Application to HTM}
The infinitely long range IM which we call HTM is frequently being
used to illustrate general concepts of phase transition theory,
such as the critical behavior, using as an example a simple
exactly solvable case \cite{salinas_introduction_2001}. Besides,
because the exact solution exhibits MF behavior, the equilibrium
HTM clarifies the physics underlying MF approximations
widely used in the analysis of various models (see, e.g., Refs.\
\cite{gouyet_description_2003,nastar2005,aguilera_unifying_2021}). A major
result of the present study is that the kinetic HTM evolving under the
Glauber dynamics can be also exactly solved in the thermodynamic limit,
as will be shown below.
\subsection{HTM at thermal equilibrium}
In HTM all spin pairs interact with the same strength $ -J/N $ so the
dimensionless Hamiltonian reads
\begin{equation}\label{HTM0}
	{\cal H}^0 =-\frac{\beta J}{2N}\sum_{i,j=1}^{N} 
	\sigma_i\sigma_j-\beta H\sum_{i=1}^{N} \sigma_i
\end{equation}
where summations over $i,j$ are not restricted to nn pairs as in the
conventional IM. Because in this case $K(\vec{e})={\beta J}/{N}$ does
not dependent on $\vec{e}$, the sum in Eq.\  (\ref{abs}) diverges as
$O(N)$. However, the divergence is compensated by the factor $1/N $
in $K$ so HTM formally satisfies the locality Eq.\  (\ref{abs}) despite
being long range.

The extensivity of HTM Hamiltonian can be easily seen if represented 
in terms of the total spin density $m$ \cite{mori_asymptotic_2010}
\begin{equation}
	M = \sum_{i=1}^N \sigma_i\equiv Nm
	\label{M}
\end{equation}
as
\begin{eqnarray}
	{\cal H}^0 &=&- \frac{\beta J}{2N} M^2 - \beta {H}M\nonumber\\
	&=&-N\left[\frac{\beta}{2}m^2+hm\right]\equiv Nu^0(m).
	\label{H00}
\end{eqnarray}
Here and in the following to simplify notation we choose $J$
as our energy unit $J=1$ and $h=\beta H$ is the dimensionless external
magnetic field.

\begin{figure}
	\begin{center}
		\includegraphics[viewport=148 474 473 667,scale = 0.7]{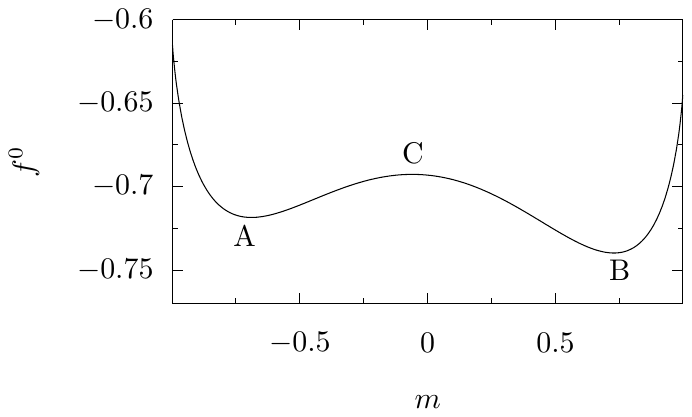}%
	\end{center}
	\caption{\label{fig1}Equilibrium FFE density Eq.\ (\ref{f})
		for $T=0.8T_c$ and $h=0.015$. A, B, and C denote,
		respectively, the two local minima and a local maximum
		appearing in $f^0(m)$ when $T<T_c$ and $|h|<|h_{SP}|$
		where $ h_{SP} $ is $ h $ at the spinodal point (further
		explanations are given in the text).}
\end{figure}

The equilibrium partition function 
\begin{equation}\label{Zeq}
	Z^{eq}=\mbox{Tr}_{\pmb{\sigma}}e^{-Nu^0(m)}=\sum_{M=-N}^{N} 
	{}_NC_{(N+M)/2}\,e^{-Nu^0(m)}
\end{equation}
where $ {}_NC_{(N+M)/2} $ is the combinatorial factor equal to the number
of combinations of $N$ spins with magnetization $M$. In the limit of
large $N$,---which will be always assumed to be the case in the present
study,---the Stirling approximation gives
\begin{equation}\label{S}
	S(N,M)=\ln {}_NC_{(N+M)/2}\simeq Ns(m)
\end{equation}
with \cite{salinas_introduction_2001,mori_asymptotic_2010}
\begin{equation}
	s(m)=-\sum_{\genfrac{(}{)}{0pt}{2}{up}{lo}}
	\frac{1\pm m}{2}\ln \frac{1\pm m}{2}.
	\label{s(m)}
\end{equation}
where the sum consists of two terms: one with all lower
signs and the other one with all upper signs.
Substituting Eq.\ (\ref{s(m)}) in Eq.\ (\ref{Zeq}) one gets
\begin{equation}\label{Zeq2}
	e^{-\beta F^{eq}(h)}\simeq \int_{-1}^1 dm e^{-\beta F^{0}(m)},
\end{equation}
where we have omitted all factors that in the thermodynamic
limit do not contribute to $f^{0}(m)$ defined as
\cite{salinas_introduction_2001,mori_asymptotic_2010}
\begin{equation}
	\beta F^0(m)\equiv Nf^0(m)=N[u^0(m)-s(m)].
	\label{f}
\end{equation}
We will call $ F^0 $ and $ f^0 $ the fluctuating free energy (FFE) and
FFE density, respectively, because unlike the true free energy they are
not convex, as can be seen in Fig.\ \ref{fig1} where a typical FFE
density for HTM below $T_c$ is drawn.
$ h_{SP} $ in Fig.\ \ref{fig1} and magnetization $ m_{SP} $ below in Eq.\
(\ref{phi}) are the values of $ h $ and $ m $ at the spinodal point
defined by conditions
\begin{equation}\label{spinodal}
f^0_m = 0\mbox{\ and\ }	f^0_{mm} = 0.
\end{equation}
Their explicit expressions can be found in Ref.\
\cite{mori_asymptotic_2010}. In Eq.\ (\ref{spinodal}) and in the following
the subscript $ m $ will denote the symmetric discrete derivative defined
for an arbitrary function $g(m)$ as
\begin{equation}
	g_m(m)=\frac{g(m+\epsilon)-g(m-\epsilon)}{2\epsilon},
	\label{dm}
\end{equation}
where $\epsilon=1/N$. 

As can be seen from Eq.\ (\ref{Zeq2}) and Fig.\ \ref{fig1}, in Laplace's
approximation suitable at large $N$ only the absolute minimum (or two
when $h=0$) will contribute to $f^{eq}$ while the contributions from the
hump in the region of $m$ between points A and B in Fig.\ \ref{fig1}
will vanish in the thermodynamic limit. However, in nonequilibrium
case the behavior of FFE in this region is of paramount importance
\cite{mori_asymptotic_2010,tome_dynamic_1990}.
\subsection{NLME for HTM}
The kinetic HTM has been studied previously within the
linear ME and the Fokker-Planck equation approaches in Refs.\
\cite{weiss-ising,matkowsky1984asymptotic,mori_asymptotic_2010}. Most
pertinent to our study will be the numerical data of Ref.\
\cite{mori_asymptotic_2010} so to facilitate comparison we will try to
closely follow the notation of that paper.

NLME for HTM can be derived straightforwardly from Eq.\ (\ref{NME})
using explicit expression for the equilibrium Hamiltonian Eq.\
(\ref{H00}), allowing the Hamiltonian parameters to depend on time,
if necessary.  However, to facilitate comparison with Ref.\
\cite{mori_asymptotic_2010}, we will derive NLME for HTM departing from
Eq.\ (7) of that reference. To this end we first note that in Ref.\
\cite{mori_asymptotic_2010} the authors considered not individual
spins $\sigma_i$ but the total magnetization $M$ as the fluctuating
quantity. Denoting its probability distribution function as $P^{(M)}$
we note that it differs from our $P$ which describes the distribution
of the fluctuating Ising spins; in our formulas $ M $ serves only as a
shorthand for the sum of spins Eq.\ (\ref{M}).  The configuration space
in our case consists of $2^N$ points while $M$ takes only $N+1$ values
from $-N$ to $N$ with step 2 (a spin flip changes $M$ by 2). This is
because many spin configurations correspond to one $M$ value which can
be accounted for by the combinatorial factor $_NC_{(N+M)/2}$ as
\begin{equation}
	P^{(M)}(M,t)=\frac{N!}{N_+!N_-!}P(M,t),
	\label{PP}
\end{equation}
where $N_+=(N+M)/2$ and $N_-=(N-M)/2$) are, respectively, the
numbers of spins in the configuration pointing up and down.

In the present notation Eq.\ (7) of Ref.\ \cite{mori_asymptotic_2010} reads
\begin{eqnarray}
&&P^{(M)}_t(M,t)=\nonumber\\
&&-\frac{1}{2\tau_0}
\sum_{\genfrac{(}{)}{0pt}{2}{up}{lo}}
\left(N_\pm\frac{\exp[\mp\beta (M\mp1)/N\mp h]}{\cosh[\beta (M\mp1)/N+h]}
P^{(M)}(M,t)\right.\nonumber\\
&&-\left.(N_\mp+1)\frac{\exp[\pm\beta (M\mp1)/N\pm h]}
{\cosh[\beta (M\mp1)/N+h]}P^{(M)}(M\mp2,t)\right),\nonumber\\
	\label{eq7-2}
\end{eqnarray}
where the summation follows the same rule as in Eq.\ (\ref{s(m)}). It
should be noted that one of two terms proportional to $P^{(M)}(M\mp2,t)$
should be set to zero at the end points $M=\pm N$ because values of
$|M|>N$ are not allowed.

Now substituting Eqs.\ (\ref{PP}) and (\ref{P-u}) into Eq.\ (\ref{eq7-2})  one
obtains NLME containing only intensive quantities :
\begin{widetext}
\begin{equation}
	u_t(m,t)=\frac{1}{4\tau_0}\sum_{\genfrac{(}{)}{0pt}{2}{up}{lo}}(1\pm 
	m)\left(
	\frac{\exp[\pm u_m^0(m\mp\epsilon,t)]}{\cosh[u_m^0(m\mp\epsilon,t)]}
	-\frac{\exp[\mp u_m^0(m\mp\epsilon,t)]}{\cosh[u_m^0(m\mp\epsilon,t)]}
	e^{\pm2 u_{{m}}(m\mp\epsilon,t)}\right).
	\label{eq7-42}
\end{equation}
\end{widetext}
As expected, each term on the rhs of Eq.\ (\ref{eq7-42}) turns to zero for
$u_m(m,t)=u_m^0(m)$ which is a direct consequence of the detailed balance
condition satisfied by $r$ in Eq.\ (\ref{rss'}). An important observation
is that Eq.\ (\ref{eq7-42}) contains only derivatives of $u$ with respect
to $t$ and $m$ so a constant contribution to EH does not change during
the evolution which makes possible normalization of NPD either in the
initial condition or at any point during the evolution. It should be
noted that no approximations have been made in the derivation, so Eq.\
(\ref{eq7-42}) is exact. Most importantly, all exponential functions
in this equation have arguments of order unity, so their computation is
unproblematic for systems of any size.
\section{\label{numerical}Decay of metastable states}
Metastable states in many-body systems emerge when there exist
local minima in the energy landscape. By definition, they should
decay as the system evolves toward thermal equilibrium. The
decay process is of paramount importance for the kinetics of phase
transitions and has been extensively investigated in a variety of systems
\cite{van1992stochastic,kramers-RMP,acharyya_m_nucleation_1998,%
Kirkpatrick1983OptimizationBS,nishimori_statistical_2001,BPZ,%
ryu_validity_2010,mori_asymptotic_2010}.
In all studies the mechanism of escape from the metastable states was
found to proceed via thermal activation with the transition rate depending
on the FFE barrier $\Delta F^*$ according to the Arrhenius law
\begin{equation}
	R\propto \exp(-\Delta F^*/k_BT),
	\label{arrhenius}
\end{equation}
where the star denotes the maximum value of the difference between
FFE along the reaction pathway and at the local minimum in which the
metastable system temporary resides.  The proportionality coefficient in
Eq.\ (\ref{arrhenius}), however, is model-dependent and in the kinetic IM
its behavior at low temperatures can be very complicated, as can be seen
from analytical calculations and simulations with the use of specialized
Monte Carlo algorithms (see Ref.\ \cite{nita} an references therein). The
latter approaches were developed to overcome the difficulty encountered
in straightforward numerical simulations at low temperatures where the
evolution develops at a prohibitively large timescale because of the
Arrhenius-law dependence on $T$.

Apparently for this reason simulations of the decay in HTM in Ref.\
\cite{mori_asymptotic_2010} were restricted to moderately low temperature
$T=0.5T_c$ and to shallow local FFE minima in the vicinity of the
spinodal point $m_{SP},h_{SP}$ defined in Eq.\ (\ref{spinodal}). The
calculated lifetimes $\tau$ of the metastable states were rather
modest,$\tau=O(10^4\tau_0)$, so to extend results to larger lifetime
values a heuristic asymptotic expression was suggested presumed to be
valid for metastable states ($h<h_{SP}$) in the limit
\begin{equation}\label{Lambda}
\vert\Lambda\vert=(h_{SP}-h)N^{2/3}\to\infty.	
\end{equation}
However, the simulations were performed only for $ \vert\Lambda\vert
\lesssim 4 $ \cite{mori_asymptotic_2010} and though good agreement with
the asymptotic expression was found, it remains unclear whether the data
are already in the asymptotic range.

At moderately low temperatures (e.g., $ \gtrsim 0.4T_c$ in 2D IM
\cite{cluster_decay}) that will be assumed also in the present paper
the qualitative behavior of the decay simplifies and phenomenological
classic nucleation theory (CNT)  \cite{nucleation} can be used for an
accurate calculation of lifetimes in the conventional kinetic IM with
nn interactions \cite{acharyya_m_nucleation_1998,ryu_validity_2010}.
CNT, however, is not suitable to deal with HTM because it heavily
relies on the short range nature of spin interactions. It is pertinent
to note that nucleation time is usually much larger than the time of
the subsequent growth of the stable phase, so in most cases the total
lifetime of a metastable state $\tau$ is dominated by the nucleation
time $\tau\simeq1/R$ which will be assumed throughout the paper.

To understand the difference between nucleation in IM and in
HTM, let us consider a ferromagnetic model with Hamiltonian Eq.\
(\ref{nnIsing}) in equilibrium ordered state below $T_c$ with, say,
negative magnetization. The Hamiltonian can describe both IM and HTM by
restricting interactions $J_{nn}$ only to nn spins in the former case
(here for clarity we temporary return to dimensional parameters) while
in the latter case the interaction is $J/N$ for any spin separation. In
both cases the decay is driven by the external field which contribution
to FFE difference in Eq.\ (\ref{arrhenius}) is negative $\sim -2Hs$,
where $s$ is the number of reversed spins within the nucleus and FFE
at low temperatures is approximated by the interaction energy. The
decay is hampered by the positive contribution due to interaction
of spins within the nucleus with oppositely directed spins in the
bulk. In IM case the nucleus can be chosen to be roughly spherical to
minimize the surface where the positive energy density is concentrated
so their contribution will scale with $s$ as $cs^{1-1/d}$ where $c$
is a size-independent positive coefficient and $d$ the space dimension
\cite{acharyya_m_nucleation_1998,ryu_validity_2010}
\begin{equation}
	\label{FIM}
\Delta F_{IM} \simeq cs^{1-1/d} -2Hs.
\end{equation}
In HTM, however, the positive contribution is not spatially localized
because of the infinite range of the interactions so any $s$ spins can be
considered as a nucleus (rather a misnomer in this case). Substituting in
Eq.\ (\ref{H00}) $M\simeq-N$ for the metastable system at low temperature
$ T\to0 $ and $M\simeq-N+2s$ for the system with the nucleus one gets
after subtraction
\begin{equation}
	\label{FHTM}
	\Delta F_{HTM}\simeq 2J(1-s/N)s - 2Hs.
\end{equation}
The maximum values of the difference needed in Eq.\ (\ref{arrhenius}) are found from the condition
\begin{equation}\label{dFds}
	d(\Delta F)/ds\vert_{s^*}=0
\end{equation}
where $ s^* $ is the size of the critical nucleus.

In the case of HTM one finds
\begin{eqnarray}\label{s*}
	s^*&=&\frac{J-H}{2J}N\\
	\label{F*}
	\Delta F_{HTM}^*&=&\frac{(J-H)^2}{2J}N.
\end{eqnarray} 
These expressions qualitatively differ from those that can be obtained
from Eq.\ (\ref{FIM}) \cite{acharyya_m_nucleation_1998,ryu_validity_2010}
in that they both linearly depend on $N$ which, in particular, means
that in thermodynamic limit the decay is impossible according to Eq.\
(\ref{arrhenius}). In the IM case, on the other hand, both quantities
remain finite as $N\to\infty$.

The absence of spatial separation between spins with different
orientations and the linear dependence of $ \Delta F_{HTM}^* $ on the
system size makes the decay in HTM qualitatively similar to the N\'eel
relaxation in single-domain magnets which will be further discussed
below in Sec.\ \ref{neel}.

One may note that Eq.\ (\ref{s*}) does not make sense for $ H>H_c=J
$. The reason is that in both HTM and IM there exist some critical
value of $H=H_c$ above which the metastability is impossible because the
necessary local minimum disappears. As can be seen from Eqs.\ (\ref{r})
and (\ref{delta_i}) the spin flip rate becomes large  $\sim\tau_0^{-1}$
irrespective of the spin configuration so spins at random positions
can flip to positive values and the system will reach equilibrium on a
microscopic time scale $O(\tau_0)$.  $H_c=J$ for HTM was obtained in the
low temperature approximation; the exact value accounting for entropy
effects is $H_{SP}$ \cite{mori_asymptotic_2010} and in IM $H_c\simeq
H_{SP}$ in MF approximation.
\subsection{\label{HTMdecay}The decay in HTM}
Because in the thermodynamic limit the decay is impossible in HTM,
it has been simulated with the use of NLME at finite $N$. In HTM the
metastable state can form when of $0\le h<h_{SP}$ in the vicinity of
minimum A in Fig.\ \ref{fig1}. This problem was investigated in Ref.\
\cite{mori_asymptotic_2010} within linear ME. Following that study the
initial condition has been chosen as the Gaussian NPD
\begin{equation}
	P^{(M)}(Nm,t=0) =\sqrt{\frac{aN}{2\pi}}
	\exp\left[-\frac{aN}{2}(m-m_A)^2\right].
	\label{Pini}
\end{equation}
The initial condition for $u$ needed in Eq.\ (\ref{eq7-42}) has been obtained
from an approximate equation
\begin{equation}
	P^{(M)}(M,t) \simeq e^{-Nf(m,t)}
	\label{P-f}
\end{equation}
where in complete analogy with Eq.\ (\ref{f})
\begin{equation}
	f(m,t)=u(m,t)-s(m).
        \label{FFE}
\end{equation}
The approximation for $u(m,t=0)$ obtained from the last three equations
consists in using Strling's formula for the combinatorial factor in
Eq.\ (\ref{PP}). This is admissible in the case of sufficiently deep
metastable minima for large $N$ that we are going to consider because
the initial condition can be arbitrary so the slight difference with
Ref.\ \cite{mori_asymptotic_2010} is not essential, as discussed in
detail below.

\begin{figure}
	\begin{center}
		\includegraphics[viewport=60 10 278 218,scale = 0.7]{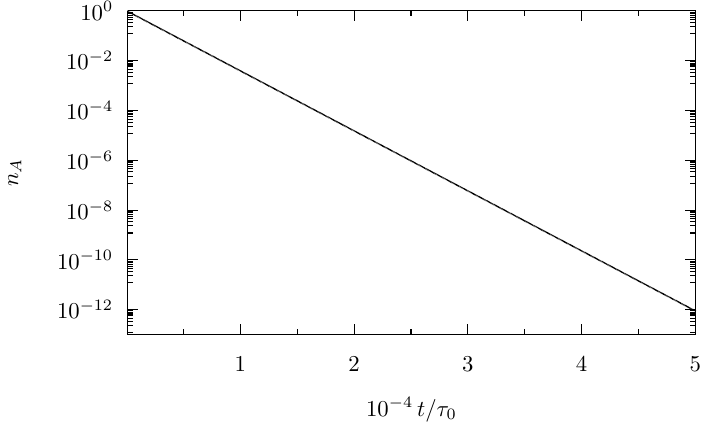}%
	\end{center}
	\caption{\label{fig2}Probability of survival of the metastable
		states Eq.\ (\ref{exp-law}) calculated with the use of
		Eq.\ (\ref{eq7-42}) for HTM with $N=10^3$, $T=0.8T_c$,
		and $h=0.06$. The points are spaced with time step
		$100\tau_0$ and their size exceeds the accuracy of
		calculations. The exponential law has been fulfilled
		better than the accuracy of the drawing (see the text).}
\end{figure}

In all calculations parameter $ a $ in Eq.\ (\ref{Pini}) has been
fixed at the same arbitrarily chosen value $ a=1 $. Obviously
that in general the initial value should strongly influence the
outcome of the evolution. However, in our  case this will not be
significant for the following reason. As was pointed out in Refs.\
\cite{weiss-ising,matkowsky1984asymptotic,mori_asymptotic_2010}, the
problem of decay of a metastable state in HTM is akin to the problem
of escape over the potential barrier in a two-well potential studied
by Kramers \cite{kramers1940}. In Ref.\ \cite{kramers-RMP} the decay
is described as follows.  At the first stage an arbitrary initial
NPD relaxes toward a local quasi-equilibrium state and on the second
stage this state slowly (in comparison with the first stage) decays
into the stable equilibrium with the magnetic momenta distributed
mainly around point B in Fig.\ \ref{fig1}. In the present study the
lifetime $ \tau $ has been determined at this second stage in contrast
to Ref.\ \cite{mori_asymptotic_2010} where the first stage was also
included. Because in our definition the properties of both the minima
and the Glauber rates  depend only on the parameters of HTM, lifetime $
\tau $ does not depend on arbitrariness of the initial distribution. Now
defining the population of metastable state as the number of systems
with negative magnetization we may describe its time evolution following
Kramers' two-state transition state theory (see Sec.\ II.C.2 in Ref.\
\cite{kramers-RMP}) as
	\begin{equation}\label{exp-law}
		n_A(t)=\int_{-1}^{0}P^{(M)}(Nm,t)dm\simeq 
		n_A^{eq}+(n_A^0-n_A^{eq})e^{-t/\tau},
	\end{equation}
where $n_A^0, n_A^{eq} $ are the populations at $t=0$ and at
thermal equilibrium, respectively.  The difference with Ref.\
\cite{mori_asymptotic_2010} is that in the present study the behavior
Eq.\ (\ref{exp-law}) has been assumed to hold only after the initial
relaxation has completed while in Ref.\ \cite{mori_asymptotic_2010}
it was considered to be approximately valid throughout the whole decay
process. In view of this difference, the comparison of our lifetime $
\tau $ with the calculations of Ref.\ \cite{mori_asymptotic_2010}
would be legitimate only if the initial fast relaxation time is
negligible in comparison with $ \tau $.  This has been achieved by
restricting consideration to sufficiently deep potential wells near
the local minimum A which can be easily satisfied in large systems
$N\geq1000 $  where the depth of even a shallow well in $ f^0 $
in Fig.\ \ref{fig1} is strongly enhanced by the factor $ N $ in the
expressions of the Arrhenius type satisfied by the lifetime in HTM
\cite{kramers1940,weiss-ising,matkowsky1984asymptotic,mori_asymptotic_2010}
\begin{equation}\label{tau}
	\tau\sim e^{N[f^0(C)-f^0(A)]}\equiv e^{N\Delta f^0}
\end{equation}
(we remind that $ \beta $ enters in $ f^0 $ as a factor). For example,
for the HTM parameters used in the calculations shown in Fig.\ \ref{fig2}
$n_A^{(eq)} \approx 10^{-38}$. Because of this, the equilibrium population
was negligible in the simulations of Ref.\ \cite{mori_asymptotic_2010} so
the authors used Eq.\ (\ref{exp-law}) with $ n_A^{eq} =0$. This has been
a good approximation also in all calculations of present paper except
at $h=0$ when $ n_A^{eq}=n_B^{eq}=1/2$ due to the symmetry of $f^0$
in this case (see Fig.\ \ref{fig1}).
The qualitative picture just described is illustrated in Fig.\ \ref{fig2}
where it can be seen that the decay law Eq.\ (\ref{exp-law}) is satisfied
at least within twelve orders of magnitude. The influence of the initial
relaxation could not be discerned at the time resolution $100\tau_0$. It
is to be noted that the lifetime $\tau$ in Fig.\ \ref{fig2} is only the
second smallest in all of our calculations so the influence of the
initial relaxation in most of them has been even weaker.

Though the decay law Eq.\ (\ref{exp-law}) is only heuristic,
in the calculations it has been satisfied with a remarkable accuracy.
The specific decay rate
\begin{equation}\label{lambda}
	\lambda=\tau^{-1}/N
\end{equation}
has been found to be equal to $5.550091955(7)\times10^{-7}$ at
$t=200\tau_0, 300\tau_0, 400\tau_0, 500\tau_0,$ and $5\cdot10^4\tau_0$,
that is, it had the same value to the accuracy in nine to ten significant
digits.

\begin{figure}
\begin{center}
\includegraphics[viewport = -10 20 484 220, scale = 0.7]{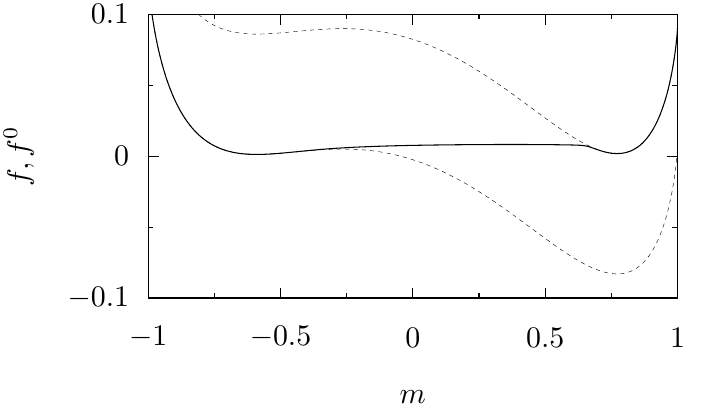}
\end{center}
\caption{\label{fig3}Solid line---FFE $ f $ corresponding to
the solution of Eq.\ (\ref{eq7-42}) at $ t=500\tau_0 $ when $ n_A \approx n_B $
with the model parameters $\beta=1.25\;(T=0.8T_c)$ and $h=0.06$ with the 
initial condition Eqs.\ (\ref{Pini})--(\ref{P-f}); dashed lines correspond to 
local fits of $ f^0 $ in Eq.\ (\ref{cacb}).}
\end{figure}

The simplicity of the behavior seen in Fig.\ \ref{fig2} suggests a
simple underlying physics which can be surmised from the behavior of
FFE during the decay shown in  Fig.\ \ref{fig3}.  As can be seen, at
intermediate times ($ t=500\tau_0 $ has been chosen as an example) $
f(m,t) $ in the vicinity of local minima A and B in Fig.\ \ref{fig1}
can be accurately approximated as
\begin{equation}\label{cacb}
	f(m,t)\approx f^0(m) + C_{A(B)}(t). 
\end{equation}
In terms of NPD this translates into two quasi-equilibrium distributions
strongly peaked near A and B and characterized by filling factors $ n_A(t) $
and $ n_B(t) =1-n_A(t)$ from Eq.\ (\ref{exp-law}).
The EH density $ u $ and FFE $ f $ differ only in time-independent
function $ s(m) $, therefore, large lifetimes of the metastable states
means a slow evolution of both.  The behavior of $ f $ in Fig.\ \ref{fig3}
near the minima is easily understood because Eq.\ (\ref{eq7-42})
has a static solution $ u=u^0+C_0 $ where the constant comes from the
fact that only derivatives  of $ u $ with respect to $ t $ and $ m $
are present in the equation. When $ u=u^0+C_0 $ holds for all $ m $, $
C_0 $ is fixed by the normalization condition as $ C_0=\ln Z$. But when
the equality is only local, as in the figure,---both minima contribute
to the normalization via $ C_A $ and $ C_B $ in Eq.\ (\ref{cacb}).

Still, the cause of the quasi-static behavior of the roughly horizontal
line that connects the two regions near the minima in Fig.\ \ref{fig3}
needs clarification. The qualitative analysis simplifies in the
thermodynamic limit where one of the parameters ($ N $) disappears
from NLME and its solutions simplify because the escape over barrier
is forbidden.
\section{\label{thrm-limit}Thermodynamic limit and the MF equation}
For our purposes in taking the limit $N\to\infty$ in NLME it is convenient,
besides setting $\epsilon=0$, to interchange in Eq.\ (\ref{eq7-42})
the second terms in the sum over the upper and the lower signs as
\begin{widetext}
\begin{equation}
	u_t(m,t)=\frac{1}{4}\sum_{\genfrac{(}{)}{0pt}{2}{up}{lo}}
	\frac{e^{\pm[u^0_m(m,t)-u_{{m}}(m,t)]}}{\cosh u^0_m(m,t) }
	\left[{e^{\pm u_{{m}}(m,t)}(1\pm m)}
	-{e^{\mp u_{{m}}(m,t)}(1\mp m)}
	\right]
	\label{eq7-5}
\end{equation}
\end{widetext}
where use has been made of the explicit form of $u^0$ Eq.\ (\ref{H00});
besides, henceforth we will choose $ \tau_0 $ as our time unit and will
omit it in evolution equations.

Now it is easily seen that the terms in large brackets in Eq.\ (\ref{eq7-5})
are nullified by
\begin{equation}\label{artanh}
	u_m = -\frac{1}{2}\ln\frac{1+m}{1-m}=-\artanh m=s_m(m)\equiv u_m^1
\end{equation}
where the penultimate equality follows from Eq.\ (\ref{s(m)}).  Thus,
there exists a locally stationary solution independent of the Hamiltonian
parameters which in terms of FFE reads
\begin{equation}\label{f1}
	f^1(m) = C_1
\end{equation}
where $ C_1 $ as a constant.
Further, after some rearrangement Eq.\ (\ref{eq7-5}) takes the form
\begin{equation}
	{u}_t=\cosh^2 u_m\, (m+\tanh u_m) (\tanh {u}^0_m-\tanh u_m)
	\label{eq1}
\end{equation}
where for brevity the arguments $(m,t)$ of all functions have been
omitted. As is seen, if $ u^0 $ is time-independent the locally stationary
solution satisfying $ u_t(m,t) =0$ is given either by $ u=u^0+Const $
due to the last factor on the rhs or by $ u^1 $ Eq.\ (\ref{artanh})
because of the second factor.

Now the structure of $ f $ seen in Fig.\ \ref{fig3} becomes qualitatively
transparent. In the thermodynamic limit the two segments of $ f$
near the minima are given by Eq.\ (\ref{cacb}) with the approximate
equality becoming exact and with  time-independent $ C_{A(B)} $, $
n_{A(B)} $ remaining unchanged with time because of the infinitely high
barrier separating the minima. At finite but sufficiently large $ N $
the solution becomes weakly time-dependent but the qualitative picture
remains accurate at large $ N $.

In the Kramers approach to escape over the potential barrier a crucial
role plays the diffusivity which is proportional to $ \epsilon =1/N$
\cite{kramers1940,weiss-ising,matkowsky1984asymptotic,mori_asymptotic_2010}
and thus can be neglected when the system is very large or
the processes studied, such as the high-frequency hysteresis
\cite{tome_dynamic_1990,chakrabarti_dynamic_1999,zhu_hysteresis_2004},
are fast in comparison with the lifetime of metastable states. In such
cases the simple Eq.\ (\ref{eq1}) should be easier to use than the system
of equations (\ref{eq7-42}) which may become unmanageable at large $ N
$. However, the method of solution of Eq.\ (\ref{eq1}) should be chosen
prudently. The problem is that numerical techniques are often based on a
finite difference approximation of the derivative over $m$ consisting
in division of interval $ m\in [-1,1] $ into, say, $L$ discretization
steps. In the case of large systems (e.g., $ N\sim 10^{23}) $ $L $ will
be much smaller than $ N $. This would effectively reduce the system size
to $ \sim L $ which may introduce a nonphysical time evolution, such as
the decay of metastable states.  A method of characteristics is devoid of
such difficulties \cite{whitham_linear_2011,kamke}. The characteristic 
equations for Eq.\ (\ref{eq1}) has been derived in Appendix 
\ref{characteristics}.
\subsection{\label{MFeq}MF equation}
In the Stirling approximation Eq.\ (\ref{PP}) can be written as
\begin{equation}\label{PP-stirling}
P^{(M)}(M,t)\simeq e^{-N[u(m,t)-s(m)]}= e^{-Nf(m,t)}.	
\end{equation}
When $ h\not=0 $ the symmetry $m\rightleftarrows-m$ is broken so in
general case there is only one global minimum in $f(m,t)$ at some $
m_0(t) $ which can be found from the condition
\begin{equation}\label{m_0}
\left.f_m(m,t)\right|_{m=m_0(t)}=\left(u_m+\artanh m\right)\vert_{m=m_0}=0
\end{equation}
where use has been made of Eqs.\ (\ref{FFE}) and (\ref{s(m)}).
For our purposes the condition is convenient to cast in the form
\begin{equation}\label{m+th}
	\left(m+\tanh u_m\right)\vert_{m=m_0}=0
\end{equation}
which in terms of the canonical variables of Appendix \ref{characteristics}
can be written as a constraint
\begin{equation}\label{constraint}
	\chi(m,q)=m+\tanh q = 0.
\end{equation}
Assuming the FFE minimum in the initial condition is at $ m_0(t_0) $ the
equation for the characteristic passing through this point is obtained
from Eqs.\ (\ref{m_dot}), (\ref{H}), (\ref{H00}), and (\ref{m+th})
as  \cite{suzuki-kubo}
\begin{equation}\label{MF}
	\dot{m}_0=-m_0+\tanh [\beta m_0 +h(t)].
\end{equation}
It is straightforward to verify that characteristic Eqs.\ (\ref{q_dot})
and (\ref{u_dot}) are also satisfied provided constraint Eq.\
(\ref{constraint}) is fulfilled along the characteristic.  This is indeed
the case because the total time derivative of $ \chi $ according to Eq.\
(\ref{g_dot}) is
\begin{equation}\label{chi_dot}
	\dot{\chi}(m,q)=\{\chi,\mathbf{H}\}\propto \chi(m,q).
\end{equation}
Thus, Eq.\ (\ref{constraint}) will be fulfilled along the characteristic
if, as we have assumed, it is satisfied at the initial point $t=t_0$.

MF Eq.\ (\ref{MF}) is a closed equation for the
average magnetization $ m_0 $ which has been widely
used in the studies of the hysteretic behavior (see, e.g.,
\cite{tome_dynamic_1990,chakrabarti_dynamic_1999,zhu_hysteresis_2004,%
carinci2013}).  In the context of the present study a useful observation
is that in the case of constant $h$ its variables can be separated and
a closed-form solution obtained.

Despite being exact, MF equation is insufficient for description of
the HTM kinetics in the thermodynamic limit.  As we saw in previous
section, the initial condition Eq.\ (\ref{Pini}) had only one extremum
but during the evolution two new extrema appeared (see Fig.\ \ref{fig3}).
This behavior cannot be described by MF Eq.\ (\ref{MF}) but is present
in Eq.\ (\ref{eq1}), as illustrates the finite-$ N $ example.  In the
thermodynamic limit such a behavior can be illustrated by the problem of
coarsening in a binary alloy \cite{coarsening}. In symmetric ($ h=0 $)
supercritical ($ T>T_c $) phase $ f^0 $ has one extremum---the minimum
at $ m_0=0 $. If quenched to a subcritical temperature $ T<T_c $ the
system will evolve toward the equilibrium state with $ f^0$ having a
double-well structure with two symmetric minima at A and B in Fig.\
\ref{fig1} with $ h=0 $. This evolution can be described by Eq.\
(\ref{eq1}) while MF equation will remain stuck at point C which will
turn into a local maximum at the end of the evolution.

Further, as is known, in MF approximation correlations between the
fluctuating variables are neglected, the average of the variables
product being approximated by the product of average values. However, the
correlations carry important information about the system. For example,
by taking statistical average of IM Hamiltonian Eq.\ (\ref{nnIsing}) it
is seen that the internal energy can be expressed exactly in terms of the
pair correlation function and the spin average. At thermal equilibrium
the pair correlation function is directly related to another important
characteristic---the magnetic susceptibility.

In kinetic HTM all correlations between the spin variables are contained in the 
moments of the magnetization 
\begin{equation}\label{momenta}
	\langle m^n(t)\rangle=\int_{-1}^1 m^nP^{(M)}(Nm,t)dm.
\end{equation}
$m_0(t)$ entering MF Eq.\ (\ref{MF}) corresponds to moment $n=1$, the pair
correlation is contained in $ \langle m^2\rangle $. As can be seen from
Eq.\ (\ref{Pini}), the latter essentially depends on $a$ at least at the
early stage of evolution while this parameter is absent in MF equation
Eq.\ (\ref{MF}).  Thus,  a complete description of the kinetic HTM can
be reached only with the use of FFE $f(m,t)$ satisfying Eq.\ (\ref{eq1}).
\section{\label{calculation}Lifetimes of metastable states}
\begin{figure}
\begin{center}
\includegraphics[viewport= -20 20 484 250,scale = 0.8]{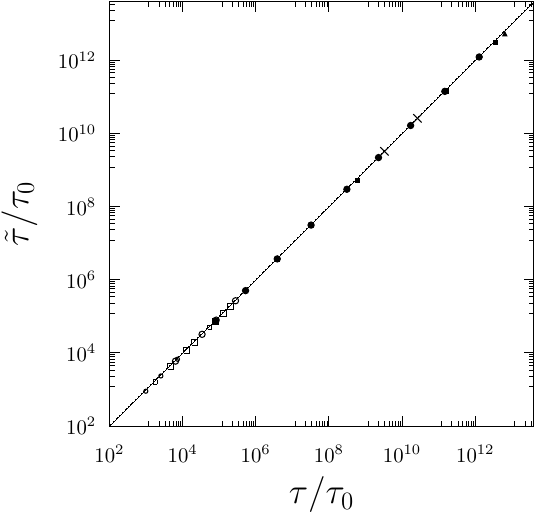}%
\end{center} 
\caption{\label{fig4}Comparison of Eq.\ (\ref{phi})
\cite{mori_asymptotic_2010} with lifetimes found from the solution of
Eqs.\ (\ref{eq7-42}) (open symbols) and (\ref{iter0}) (filled symbols and
crosses). Circles correspond to $T=0.5T_c$, squares to $T=0.8T_c$; smaller
(larger) symbols correspond to $N=10^3$ ($N=2\cdot10^3$). The cross at
larger (smaller) $\tau$ is for $N=10^5$ ($N=10^6$); the triangle marks
symmetric case $h=0$ with $N=10^3$ and $T=0.8T_c$; the straight line
corresponds to $\tilde{\tau} = \tau$. The symbol sizes do not reflect
the data accuracy which is higher than the drawing resolution.}
\end{figure}

Straightforward calculation of lifetime $ \tau $ via solution of
NLME Eq.\ (\ref{eq7-42}) and the fit to Eq.\ (\ref{exp-law}) has been
performed for HTM of two sizes $ N=10^3 $ and $ 2\cdot 10^3$, two
temperatures $ T=0.5T_c $ and $ 0.8T_c $, and variety of $ 0\le h < h_{SP}$
values. The results are shown in Fig.\ \ref{fig4} by open symbols. Larger
system sizes were not used because to achieve higher accuracy in Eq.\
(\ref{exp-law}) the integration over $ m $ has been replaced by the sum
over the discrete values 
\begin{equation}\label{m-n}
	m=-1+2n\epsilon, \quad n=0,1,2,\dots
\end{equation} 
to use the exact combinatorial factor in $
P^{(M)} $ Eq.\ (\ref{PP}). But this necessitated calculation of $ N! $
which has been found to be numerically difficult for $ N > 2\cdot10^3$ 
cases studied in Ref.\ \cite{mori_asymptotic_2010}. The use of the
Stirling approximation would be sufficient from practical standpoint
but to substantiate by numerical arguments a heuristic technique developed 
below in Sec.\ \ref{recurrence}, high precision calculations were necessary.
Though the use of NLME has made possible to extend the range
of calculated lifetimes on an order of magnitude in comparison with Ref.\
\cite{mori_asymptotic_2010}, the simulation time grew very quickly
with $ \tau $ so determination of much larger values would have required
prohibitively long calculations.  Therefore, more efficient, though heuristic
technique has been developed.  
\subsection{\label{recurrence}Recurrence relation} 
\begin{figure}
	\begin{center}
		\includegraphics[viewport=10 20 310 200,scale = 0.72]{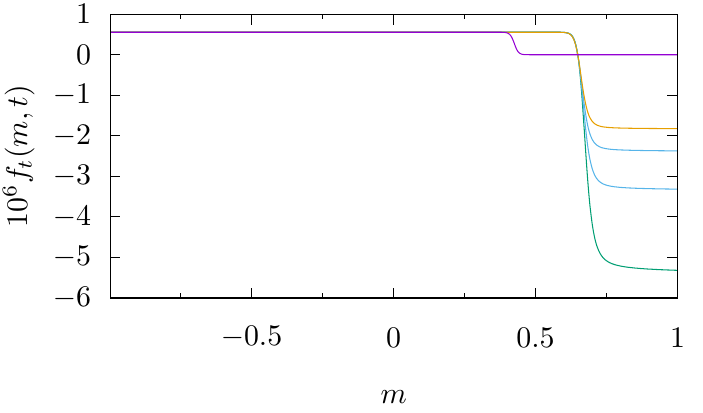}%
	\end{center} 
	\caption{\label{fig5}(Color online) Time derivative of FFE found
	numerically from Eq.\ (\ref{eq7-42}) for HTM with the same
	parameters as in Fig.\ \ref{fig2} and for the same five time values of
	(from bottom to top) as in the text following Eq.\
	(\ref{lambda}).}
\end{figure}

To
clarify the origin of the exponential law Eq.\ (\ref{exp-law}) in NLME
solutions, let us consider the evolution of FFE shown in Fig.\ \ref{fig3}
in more detail. We first note that because the configurational entropy
Eq.\ (\ref{s(m)}) does not depend on time, the time derivative of EH in
Eq.\ (\ref{eq7-42}) is equal to the time derivative of FFE so using the
available solution for $u$ the derivative could be calculated numerically,
as shown in Fig.\ \ref{fig5}.
As is seen,  the time derivative is constant in the region $m\le0 $
needed in Eq.\ (\ref{exp-law}) at all simulated times. Furthermore, it
is easy to see that in this range it should be equal to specific rate $
\lambda $ from Eq.\ (\ref{lambda}) which means that the distribution of $
m $ remains the same throughout the evolution and only the total density
of metastable states changes with time according to the exponential law
Eq.\ (\ref{exp-law}). In fact, because the evolution is very slow it is
reasonable to expect that the magnetization distribution should be close
to the equilibrium one Eq.\ (\ref{equilibrium}) as illustrated in Fig.\
\ref{fig3}. In other words, at negative values of $ m $ an accurate
solution to EH density should be feasible with the {\emph{ansatz}}
\begin{equation}\label{metastable}
	u(m,t)\simeq\lambda t + v(m,\lambda).
\end{equation} 
After substitution in Eq.\ (\ref{eq7-42}) one sees that
the time variable disappears from the equation. Also, the two terms in
the sum would contain $ v_m $ at two successive points $ m-\epsilon $
and $ m+\epsilon $, so if the leftmost $ m=-1 $ value is known, the rest
can be found successively by recursion \cite{recurrence}.  But at $
m=-1 $ only one term remains on the rhs of Eq.\ (\ref{eq7-42}), so $
v(m=-1+\epsilon) $ can be expressed through $\lambda$ and the HTM
parameters. Next introducing 
\begin{equation}\label{x-def}
	x_{n+1}=e^{-2[v_m(m+\epsilon)-u^0_m(m+\epsilon)]}-1,
\end{equation} 
Eq.\ (\ref{eq7-42}) for $x_n$ can be cast in the form of
a nonlinear recurrence relation 
\begin{equation}\label{iter0}
	x_{n+1}=a_n\frac{x_{n}}{1+x_{n}}+b_n
\end{equation} 
where 
\begin{eqnarray}
	\label{a_n}
&&a_n=\frac{1+m}{1-m}e^{2u_m^0(m+\epsilon)}
\frac{1+\exp[-2u_m^0(m+\epsilon)]}{1+\exp[-2u_m^0(m-\epsilon)]}\\
&&\mbox{and}\nonumber\\
\label{b_n}
&&b_n=-\frac{2\lambda}{1-m}\left(1+e^{2u_m^0(m+\epsilon)}\right).
\end{eqnarray} 
(Note that in the above equations only subscript $ m $ stands for the
discrete derivative, subscripts $ n,\;n+1 $ are just integer numbers.)
Thus, from the recurrence relation Eq.\ (\ref{iter0}) all $ v(m,\lambda)
$ can be found provided a suitable value of $ \lambda $ is chosen in Eq.\
(\ref{b_n}). To understand how this can be done, a more careful analysis
of Eqs.\ (\ref{iter0})--(\ref{b_n}) is in order.

To this end we first introduce a useful formal rearrangement of the
recurrence. As can be seen from Eq.\ (\ref{a_n}), $a_0$ is equal to zero
so that according to Eq.\ (\ref{iter0}) $ x_1 $ is equal to $ b_0 $. Thus,
the recurrence can be initiated by $ x_1 $ which, as can be seen from
Eq.\ (\ref{m-n}), has the lowest physically allowed subscript. However,
a simpler solution of the linear recurrence in Appendix \ref{linear}
is obtained if it is initiated by $ x_0=0 $ and $ a_0=1 $. This is
admissible because for $ m=-1 $ ($ n=0 $) the first term on the rhs of
Eq.\ (\ref{iter0}) vanishes, so with $ x_0=0 $ $a_0$ may be ascribed
any value.

\begin{figure}
\begin{center}
	\includegraphics[viewport=-10 20 484 220,scale = 0.7]{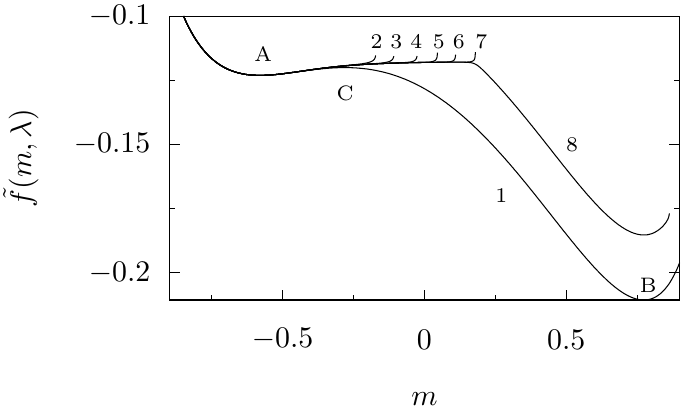}%
\end{center} 
\caption{\label{fig6}FFE Eq.\ (\ref{ffe-iter}) calculated with the use
of Eqs.\ (\ref{iter0}) and (\ref{x-def}) for values of $\lambda$ given
in Table \ref{table}.}
\end{figure}

A trivial solution of the recurrence is obtained with the choice $
\lambda =0$. Then from Eqs.\ (\ref{iter0}) and (\ref{b_n}) follows that
$x_n=0$. According to Eq.\ (\ref{x-def}) this means that $ v_m=u^0_m $,
that is, the equilibrium solution is recovered. This ought to be expected
because $ \lambda =0$ corresponds to a stationary state.  Solutions of
Eq.\ (\ref{iter0}) will be convenient to visualize with the use of FFE
\begin{equation}\label{ffe-iter}
	\tilde{f}(m,\lambda)=v(m,\lambda)-s(m).
\end{equation} Curve 1 in Fig.\ \ref{fig6} corresponds to the fully
decayed state (barring the remainder $n_A\sim 10^{-38} $), i.e., to the
equilibrium state $ v=u^0 $.  
	\begin{table}[h] \begin{center}
		\caption{Values of parameter $\lambda$ for curves shown
		in Fig.\ \ref{fig6} and the
			fit of data in Fig.\ \ref{fig2} to Eq.\
			(\ref{exp-law}).}
		\label{table} \begin{tabular}{c|l}
			Curve&\mbox{\hspace{2em}$10^{7}\lambda$}\\
			\hline 1&0.0\\ 2&6.0\\ 3&5.6\\ 4&5.551\\
			5&5.5501\\ 6&5.550092\\ 7&5.5500919602\\
			8&5.5500919601\\ \hline Fit&5.55009
		\end{tabular}
	\end{center}
\end{table} 
A less trivial solution can be obtained for small $ x_n $
satisfying \begin{equation}\label{x-small}
	\max|x_n|\ll 1
\end{equation} 
in which case $ x_n $ in the denominator in Eq.\
(\ref{iter0}) can be neglected and a closed-form solution obtained
(see Appendix \ref{linear}) 
\begin{equation}\label{x-linear}
	x_n=\sum_{l=0}^{n-1}b_l \exp\left(\sum_{k=l+1\leq n-1}^{n-1}\ln
	a_k\right).
\end{equation} 
Using Eq.\ (\ref{b_n}) two immediate conclusions can be made:
(i) $ x_n\propto\lambda $ and (ii)  all $ x_n$ for $ n\geq 1 $ are
negative. Thus, when $ \max|x_n|\to1$ ($ x_n\to-1 $) a singularity at $
m=-1+2n\epsilon $ corresponding to $ n $ will arise in $ x_n $, hence,
also in EH and FFE Eq.\ (\ref{ffe-iter}) which means that the solution
is unphysical at this value of $ \lambda $.

Thus, the physical values should be sought in the interval $ 0\leq
\lambda\leq\lambda_{max} $ with the upper limit found as the largest
$ \lambda $ at which the solution does not diverge. This has been done
via a trial and error process for the same HTM parameters as in Fig.\
\ref{fig2}. The results are presented in Fig.\ \ref{fig6} and Table
\ref{table}. The range of calculations has been extended beyond $
m=0 $ into the region where according to Fig.\ \ref{fig5} solution Eq.\
(\ref{metastable}) is still valid. It has been found that $ v(m,\lambda) $
is very sensitive to the precise value of $ \lambda $ near $ \lambda_{max}
$. For example, curves 7 and 8 correspond to $ \lambda $ values that
differ on a unit in the eleventh significant digit. Yet curve 7 shows that
the value is still too large (the solution diverges) while curve 8 already
goes deep down and if assumed to be correct predicts $ n_A\sim 10^{-27}
$, that is the decay is practically terminated. In fact, the curve is
drawn farther than the range of validity of Eq.\ (\ref{iter0}) as can
be seen by comparing Figs.\ \ref{fig5} and \ref{fig6}. But even at the
upper end of the region of validity $ m\lesssim0.4 $ the value of $ \tilde{f}
$ is such that $ n_A\sim10^{-10} $. This, however, is an overestimation
because the negative derivative of curve 8 indicates that the true value
of $ \tilde{f} $ near minimum B should be noticeably smaller. Anyway,
we are going to compare the calculated lifetimes with those of Ref.\
\cite{mori_asymptotic_2010} defined as the weighted time averages
so the contributions of small $ n_A(t) $ were insignificant. Thus,
in our case a single value of $ \lambda_{max}$ should be sufficient to
determine the lifetime from Eq.\ (\ref{lambda}). This is consistent with
the data shown in Figs.\ \ref{fig2} and \ref{fig5} where $ \lambda\approx
5.55009\cdot10^{-7} $ (see Table \ref{table}) was sufficient to describe
the decay from $ n_A\approx0.9 $ to $ 10^{-12} $. The agreement of $
\lambda $ found from the recurrence relation with that determined in a
fit to the solution of the exact evolution equation to all six significant
digits provided by the fitting software supports the suggestion that
the heuristic technique based on recurrence Eq.\ (\ref{iter0}) does
make possible an accurate determination of lifetimes of the metastable
states in HTM. Similar agreement between the two techniques has
been found for several other sets of HTM parameters.

The values of $ \tau $ calculated in this way are shown in Fig.\
\ref{fig4} by filled symbols. The excellent agreement with the results
of Ref.\ \cite{mori_asymptotic_2010} is illustrated by comparison
with the analytic expression suggested in that paper 
\begin{equation}\label{phi}
	\tilde{\tau}=\frac{\pi}{\sqrt{|m_{SP}|(h_{SP}-h)}}e^{N\Delta f^0}.
\end{equation} 
where $ \Delta f^0 $ has been defined in Eq.\
(\ref{tau}). In our calculations the notion of scaling in the vicinity
of the spinodal has not been used because in some cases, such as the one
depicted in Fig.\ \ref{fig1}, the simulated systems were quite far
from it. Therefore, to make comparison with Eq.\ (41) of Ref.\
\cite{mori_asymptotic_2010} $ \Delta f^0 $ for the barrier height has
been used instead of the scaling expression. The comparison with the
latter has also been performed with the agreement being only slightly
worse than in Fig.\ \ref{fig4}, arguably, for the above mentioned
reason. The upper limit of the data presented in Fig.\ \ref{fig4}
has been defined by the fact that for the calculations to be
meaningful the second term in the denominator of Eq.\ (\ref{iter0}) $
1+x_n $ should contribute to the recursion from $ n=1 $ onward. Therefore,
$ n_1=O(\lambda) $ should exceed the smallest number able to change the
result when added to unity. In the double precision arithmetic used in
calculations it should be larger than $ \sim 2\times10^{-16}$. The use
of software with this quantity being much smaller would make possible
to predict much larger lifetimes.

In principle, analytical expression of the type of Eq.\ (\ref{phi})
should be possible to derive from the condition $x_n \to-1$.  However,
this would necessitate knowledge of an analytical expression for $
x_n $ which would be difficult to obtain because Eq.\ (\ref{iter0})
becomes strongly nonlinear in this regime.

However, the leading exponential behavior in $ N $  can be estimated
in the linear approximation Eq.\ (\ref{x-linear}).  To this end in
the expression Eq.\ (\ref{b_n}) for $b_n$ we retain only factor $
\lambda $, which needs to be found, and in the summation over $k$
only the first two factors in the expression Eq.\ (\ref{a_n}) will be
kept because the last factor does not contribute to the logarithm in
the limit $ \epsilon\to0 $. With the use of Eq.\ (\ref{s(m)}) the first
two factors in Eq.\ (\ref{a_n}) can be unified in $\exp[2 f^0_m(m_k)]$;
here and below $ m_n,m_l$ and $ m_k$ are connected to $ n,l,k $  via Eq.\
(\ref{m-n}). Next, approximating $2\sum_{l(k)}\approx N\int dm_{l(k)}$
($ dm_{l(k)} \simeq 2/N $) and integrating over $ dm_k $ one arrives at
an estimate 
\begin{eqnarray}\label{x-approx}
	\max_n|x_n|&\sim& \max_{m_n}\lambda e^{Nf^0(m_n)}\int_{-1}^{m_n}
	dm_l e^{-Nf^0(m_l)}\nonumber\\ &\sim&\lambda
	e^{N[f^0(C)-f^0(A)]}\equiv\lambda e^{N\Delta f^0}\simeq1.
\end{eqnarray} 
where the integration over $ m_l $ for
large $N$ has been estimated by Laplace's method, and within the range
of validity of Eq.\ (\ref{iter0}) the maximum is attained at $m_n= m_C $.
The estimate Eq.\ (\ref{tau}) now follows from Eq.\ (\ref{lambda}).

The evolution of HTM for $ h>h_{SP} $ (in the geometry of Fig.\
\ref{fig1}) has been found to be not very interesting, mainly because it
is not universal as there is no locally stable minimum in $ f^0 $ at $
m_A $. The decay of initial state Eq.\ (\ref{Pini}) starts immediately and
strongly depends on the arbitrary parameter $a$. At large $ h\gg h_{SP} $
the minimum in $ f^0 $ will be moving toward $ m_B $  according to MF Eq.\
(\ref{MF}) in which case the time dependence of $ n_A $ will be close
to the step function $ n_A(t)\approx\theta[-m_0(t)] $. However, in the
vicinity of the spinodal when $ h$ approaches $ h_{SP} $ from above the
MF evolution slows down \cite{mori_asymptotic_2010} and the diffusion
mechanism starts to dominate. The exponential behavior similar to Eq.\
(\ref{exp-law}) has been observed for $ h\to h_{SP}^+ $ but in the absence
of the double-well structure of $ f^0 $ its origin is obscure.  The second
minimum at $ m_B $ has been formed which presupposes the existence of
the barrier in FFE $ f $ $h>h_{SP}$  but whether this kinetically induced
shape plays the same role as the double-well structure of equilibrium $
f^0 $ is unclear. Besides, when $h>h_{SP}$ the evolution towards thermal
equilibrium has been found to be very fast so analytical interpolation
to numerically inaccessible regions is not as useful as in the case of
escape over high barriers.

To sum up, the techniques developed in the present paper have made
possible to substantiate the heuristic expression Eq.\ (\ref{phi})
for the lifetime of metastable states in HTM suggested in Ref.\
\cite{mori_asymptotic_2010} for about eight orders of magnitude larger
values of $\tau$, an order of magnitude larger $ \vert\Lambda\vert >
20 $ in Eq.\ (\ref{Lambda}), and for FFE wells farther from the spinodal
point up to the symmetric case at zero external field corresponding to
the deepest possible well at a given temperature.
\subsection{\label{neel}N\'eel relaxation and hysteresis}
As has been noted earlier, the absence of surfaces separating the
regions of differently oriented spins,---hence, no mechanism of
domain formation,---as well as the exponential dependence of the
metastable lifetimes on the system size makes HTM a viable model
of the N\'eel relaxation (NR) \cite{neel1949} in single-domain
magnets with strong uniaxial anisotropy. Additional argument in
favor of this possibility gives a common practice to describe NR
within the framework of Brown's discrete orientation model based on
the Kramers transition-state theory of escape from the potential
well \cite{neel_relaxation_review,neel_relaxation,kramers-RMP}
which also successfully describes HTM data, as we saw in Sec.\
\ref{HTMdecay}.  An advantage of the HTM approach is that unlike in
Brown's description the barrier separating local minima need not be
high \cite{neel_relaxation_review,neel_relaxation}.  The restriction can
be detrimental in modeling hysteresis with the external pumping field
amplitude exceeding $H_{SP}$ when the energy well becomes shallow or
disappears altogether during some time intervals.

Another feature of hysteresis in single-domain magnets that has not
yet been satisfactorily described theoretically is that the area
of the hysteresis loop should tend to zero in the zero-frequency
limit \cite{chakrabarti_dynamic_1999,2DIsing1998}. However,
this behavior is not reproduced by MF equations
which has been frequently used in hysteresis studies
\cite{tome_dynamic_1990,chakrabarti_dynamic_1999,zhu_hysteresis_2004}. The
MF equation predicts the loop area to be finite at zero frequency which
could be attributed to the inadequacy of MF approximation in slowly
varying external fields. But as we have shown in Sec.\ \ref{MFeq},
in the thermodynamic limit the average magnetization in HTM satisfies
the MF equation exactly which may cast doubt on the soundness of HTM
in the description of hysteresis.  Apparently, the explanation lies in
the fact that in nonequilibrium kinetics the limits $ N\to\infty $ and $
t\to\infty $ are not always interchangeable \cite{foerster}. In HTM this
can be seen from Eq.\ (\ref{exp-law}) where the limits $\tau\to\infty$
and $ t\to\infty $ taken in different order give different results. This
problem does not arise in short range IM where $\tau$ is finite.

Because the equilibrium can be reached only if $\tau$ is finite, in
HTM this means finite $N$. The characteristic time scale separating the
two types of behavior is determined by $\tau(h=0)$ which in the present
context can be identified with NR time $ \tau_{\textrm{N}} $. When the
period $ 1/\nu $ ($ \nu $ is the oscillation frequency) of the  external
field oscillations is much smaller then $\tau$, the escape over barrier
contributes negligibly to the evolution so hysteresis can be described
within the MF approach. In the opposite limit $ 1/\nu\gg\tau $ the
system equilibrates by means of escapes. At $ h=0 $ the magnetization
will be close to zero, as can be seen from Eq.\ (\ref{exp-law}) where
at $t\to\infty$ $ n_A,n_B\to1/2 $ the densities of up and down spins
will be the same. The response to a weak external field in this case
will be that of a paramagnet or, more precisely, of a superparamegnet
because the magnetized nanoparticles are not elementary spins and the
temperature is below $T_c$. In  the $ 1/\nu\gg\tau $ and weak external
field the magnetic properties can be phenomenologically described within
the linear response theory  \cite{foerster,ROSENSWEIG2002370}.

In recent years, hysteresis and NR in single-domain magnetic
nanoparticles attracted much attention in connection with biomedical
applications, in particular, in hyperthermia via hysteresis losses
\cite{biomed,Hergt_2008,Ruta2015}. This interdisciplinary technique
has many aspects that need be investigated to develop its
comprehensive description. The results of the present paper can be useful
in qualitatively clarifying a difficulty encountered in the studies of
hysteresis which was pointed out in Ref.\ \cite{Ruta2015}. Namely, while
purely hysteretic behavior for $\tau \gg \nu^{-1}$ and superparamagnetic
behavior for $\tau \ll \nu^{-1}$ are covered by existing phenomenological
approaches, though only for small amplitudes of the oscillating
field in the second case \cite{ROSENSWEIG2002370}, the intermediate
regime $\tau \sim \nu^{-1}$ has not yet been satisfactorily described
theoretically. But it cannot be avoided in realistic setups because the
admissible for biomedical applications nanoparticles have broad size
distributions so that the whole region of $\tau(N)$ values may be covered
in practice. In the HTM-based approach all frequencies and all field
amplitudes can be described within the same formalism, though the problem
of very long simulation times needed at large $N$ should be addressed.
\section{\label{conclusion}Conclusion}
In this paper a nonlinear master equation (NLME) describing the evolution
of the effective Hamiltonian (EH) density has been suggested to overcome
the numerical difficulties caused by the exponential dependence of
nonequilibrium probability distribution (NPD) that enter into the linear
ME \cite{van1992stochastic} on the system size  $ N $.  In contrast,
NLME scales at most as $ O(N) $ and can be reduced to a set of $O(1)$
equations in the case of translationally invariant systems.

To illustrate some salient features of NLME in a simple framework,
the problem of decay of metastable states in the kinetic
Husimi-Temperley model (HTM) has been considered.  The problem
was previously studied in the framework of linear ME in Refs.\
\cite{weiss-ising,matkowsky1984asymptotic,mori_asymptotic_2010} which
results have been used for comparison purposes. An excellent agreement has
been found between numerical NLME solution and the asymptotic analytic
expression for the lifetime of metastable states suggested in Ref.\
\cite{mori_asymptotic_2010} for decay in the vicinity of the spinodal.
With the use of NLME it has been possible to cover much broader range
of parameters and to achieve much better accuracy. In particular, far
from spinodal case of zero external field has been simulated.

The exponential dependence of lifetime on the system size in HTM
ensures that in macroscopic systems the lifetime will reach values so
large that from physical standpoint the metastable states will behave
as effectively stable ones. Because of this, for large $ N $ it is
reasonable to take thermodynamic limit $ N \to\infty $. It has been
shown that in this case NLME simplifies to a nonlinear first-order
differential equation possessing, in particular, two locally stable
solutions which can be combined to construct stationary EH different
from the equilibrium one. To solve the differential equation, a system of
characteristic equations has been derived which, in particular, reduces
to the conventional MF equation \cite{suzuki-kubo} for magnetization
corresponding to the fluctuating free energy extrema.

The MF equation that has been widely used in modeling hysteresis
\cite{tome_dynamic_1990,chakrabarti_dynamic_1999} was found to fail
in the low frequency region \cite{chakrabarti_dynamic_1999}. In the
present paper it has been shown that at large but finite system size NLME
should be able to qualitatively reproduce the correct behavior. Besides
purely theoretical interest this should also be useful in modeling
N\'eel relaxation in magnetic nanoparticles which is important in some
biomedical applications \cite{biomed}.

An example of NLME application to IM was discussed in Ref.\
\cite{PRE96}. In a simple pair approximation to EH (which was
also used in Ref.\ \cite{nastar2005}) it was found that NLME
leads to a qualitatively more sound description of the spinodal
decomposition than the MF approximation to ME used, e.g., in Ref.\
\cite{gouyet_description_2003}. Specifically, NLME predicted a power-law
growth of the volumes of the separating phases while the MF approximation
predicts an exponential behavior. The latter is incompatible with the
relaxational nature of the stochastic dynamics where the growth exponents
cannot be positive \cite{van1992stochastic}. Besides, the characteristic
length scale in the MF solution remains constant throughout the growth
while NLME predicts a coarse-graining behavior in qualitative agreement
with experimental observations.

Switching from a linear ME to the nonlinear equation may seem
counterintuitive because the former can be amenable to treatment with
the use of powerful techniques of liner algebra.  However, they seems
to be efficient only when the stochastic matrices are much smaller than
the Avogadro number characteristic of the size of macroscopic systems
\cite{finite-chains}. In statistical physics this case is usually studied
in the thermodynamic limit. But NLME has been obtained via substitution
of EH into the Boltzmann factor.  The subsequent thermodynamic limit has
led to a well defined NLME for EH which, however, cannot be linearized
back because the corresponding NPD would contain the Boltzmann factor
with infinite argument.

The inherently nonlinear nature of ME was also noted in Ref.\
\cite{van1992stochastic} (see Remark in Ch.\ V.8) where it was argued that
a linear/nonlinear dichotomy is purely mathematical and does not reflect
the underlying physics.  For example, the linear Liouville equation
is equivalent to Newton's equations which in the case of interacting
particles are nonlinear, yet it is the latter that are used in most
calculations. Yet another example is given by the BBGKY hierarchy of an
infinite number of linear equations which in practical calculations is
usually approximated by a finite number of nonlinear ones, such as the
Boltzmann transport equation.

These observations may signify that stochastic kinetics in the
thermodynamic limit is inherently nonlinear.
\begin{acknowledgments}
	I express my gratitude to Hugues Dreyss\'e for
	his hospitality, encouragement, and interest in the work. I
	am grateful to Universit\'e de Strasbourg and IPCMS for 
	support.
\end{acknowledgments}
\appendix
\section{\label{characteristics}The Hamilton-Jacobi formalism }
Because Eq.\ (\ref{eq1}) contains only derivatives of $u$ but not the
function itself, according to Ref.\ \cite{kamke} it can be cast in the
form of the Hamilton–Jacobi equation \cite{h-j}
\begin{equation}
	u_t+ \mathbf{H}(m,q,t)=0,
	\label{H-J}
\end{equation}
where
\begin{equation}
	q=u_m(m,t) \label{q_def}
\end{equation} 
and 
\begin{equation}
	\mathbf{H} = - \cosh^2 q\, (m+\tanh q) [\tanh {u}^0_m(m,t)-\tanh q]
	\label{H}
\end{equation}
is the rhs of Eq.\ (\ref{eq1}) with minus sign. In the Hamiltonian
formalism the total time derivative of any function $g(m,q,t)$ of the
``coordinate'' $m$ and ``momentum'' $q$ can be calculated as
\begin{equation}
	\dot{g}\equiv\frac{dg }{dt} = \frac{\partial g}{ \partial t} + 
	\{g, \mathbf{H}\}, \label{g_dot}
\end{equation} 
where the Poisson bracket is defined as $\{a,b\}=a_mb_q-a_qb_m$.
Now the canonical Hamiltonian equations are easily found as
\begin{eqnarray}
	\label{m_dot}
	\dot{m}&=&\mathbf{H}_q\\
	\label{q_dot}
	\dot{q}&=&-\mathbf{H}_m.
\end{eqnarray}
Further assuming that at some $t=t_0$ an initial Hamiltonian density
$u(m,t_0)$ is known, by choosing any admissible value $m_0$ and setting
$q(t_0)$ equal to $u_m(m_0,t_0)$ one can find $m(t)$ and $q(t)$ from
the solution of the initial value problem for Eqs.\ (\ref{m_dot}) and
(\ref{q_dot}). Next integrating equation
\begin{equation}
	\dot{u}=q\mathbf{H}_q-\mathbf{H}\label{u_dot}
\end{equation}
obtained from Eqs.\ (\ref{g_dot}), (\ref{m_dot}), (\ref{q_def}), and
(\ref{H-J}) one arrives at a solution for $ u(m,t) $ in parametric
form where at each $t$ the values of $u$ at different $m$ are found
from the solution of the above initial-value problem for all possible
$m_0=m(t_0)$. Such a solution is a particular case of the general method
of characteristics \cite{whitham_linear_2011}.

A rigorous derivation of the Hamiltonian-Jacobi formalism in the general
case of many variables can be found in Ref.\ \cite{kamke} but for our Eq.\
(\ref{H-J}) the following heuristic arguments should suffice. First we
observe that the partial derivative $u_m$ in Eq.\ (\ref{eq1}) couples
the values of function $u(m,t)$ at neighbor points $ m \pm\epsilon$ so by
using, e.g., the method of lines one should solve a system of $N\to\infty$
coupled ordinary evolution equations. In the method of characteristics
one tries to reduce the problem to only a few such equations. To this
end one may observe that by taking the total time derivative of Eq.\
(\ref{q_def}) in the conventional way one gets
\begin{equation}\label{dot_q}
	\dot{q}=u_{mt}+u_{mm}\dot{m}=-\mathbf{H}_m-\mathbf{H}_qu_{mm}+u_{mm}\dot{m}	
\end{equation}
where the second equality has been obtained by differentiating Eq.\
(\ref{H-J}) with respect to $ m $ by considering $q$ in $ \mathbf{H} $ as just
another notation for $ u_m $. Now if we \emph{demand} that Eq.\
(\ref{m_dot}) was satisfied, then Eq.\ (\ref{q_dot}) would follow
from Eq.\ (\ref{dot_q}). Eq.\ (\ref{u_dot}) also is easily obtained by
differentiating $ u(m,t) $, using Eq.\ (\ref{H-J} and applying the chain
rule. Next assuming $ m=m(t) $ on the characteristic line, substituting it
in Eqs.\ (\ref{m_dot}), (\ref{q_dot}), and (\ref{u_dot}) one arrives at
a system of three evolution equations for three unknown functions. It is
to be noted that Eqs.\ (\ref{m_dot}) and (\ref{q_dot}) are sufficient to
derive Eq.\ (\ref{g_dot}) which shows the consistency of our assumptions.
\section{\label{linear}Linear recurrence \cite{recurrence}}
Let us consider a linear recurrence relation 
\begin{equation}
	x_{n+1} = a_nx_n+b_n
	\label{n2n+1}
\end{equation}
initiated by $x_0$; $a_n$ and $b_n$, $ n=0,N $ are presumed to be
known. Now by introducing
\begin{equation}
	X_{n}=\frac{x_{n}}{\prod_{k=0}^{n-1}a_k}
	\label{Xn+1}
\end{equation}
and dividing both sides of Eq.\ (\ref{n2n+1}) by $\prod_{k=0}^na_k$
one arrives at a linear difference equation
\begin{equation}
	X_{n+1}-X_{n}=b_n/\prod_{k=0}^na_k
	\label{diff-eq}
\end{equation}
which can be solved as 
\begin{equation}
	X_{n}=X_0+\sum_{l=0}^{n-1}\frac{b_l}{\prod_{k=0}^la_k}.
	\label{Xsol}
\end{equation}
In the main text it was assumed that $x_0=0$ which in combination with
Eq.\ (\ref{Xn+1}) gives $ X_0=0 $ and
\begin{equation}
x_{n} = \sum_{l=0}^{n-1}b_l\prod_{k=l+1\leq n-1}^{n-1}a_k=\sum_{l=0}^{n-1}b_l
\exp\left(\sum_{k=l+1\leq n-1}^{n-1}\ln a_k\right).
\label{xn+1}
\end{equation}
\end{document}